\documentclass[12pt,twocolumns,article,nofontune]{IEEEtran}

\usepackage[acronyms,nonumberlist,nopostdot,nomain,nogroupskip]{glossaries}

\usepackage[numbers,sort&compress]{natbib}

\usepackage{graphicx}
\usepackage{tabularx}
\usepackage{lipsum}
\usepackage{amsmath}
\usepackage{wrapfig}
\usepackage{algorithmic}
\usepackage{algorithm}
\usepackage{verbatim}
\usepackage{amsmath}
\usepackage{mathtools}
\usepackage{setspace}
\usepackage[usenames]{color}
\usepackage{multirow}
\usepackage{xcolor}
\usepackage{adjustbox}
\usepackage{sidecap}
\usepackage{array}
\usepackage{hhline}
\usepackage{colortbl}% http://ctan.org/pkg/xcolor

\usepackage{amssymb}
\usepackage{amsthm}
\usepackage{xfrac}
\usepackage{color}
\usepackage{graphicx}
\usepackage{stfloats}
\usepackage{algorithm}
\usepackage{algorithmic}
\usepackage{flushend}
\usepackage{enumitem}

\usepackage{amsmath}
\usepackage{amssymb}
\usepackage[normalem]{ulem}
\usepackage{graphicx}
\usepackage{flushend}
\usepackage{url}
\usepackage{color}
\usepackage{algorithm}
\usepackage{algorithmic}
\usepackage{ragged2e}
\usepackage{amssymb}
\usepackage{fancyvrb}

\usepackage{tikz}
\usepackage{pgfplots}
\pgfplotsset{compat=newest}
\pgfplotsset{plot coordinates/math parser=false}
\newlength\fheight
\newlength\fwidth
\usetikzlibrary{plotmarks,patterns,decorations.pathreplacing,backgrounds,calc,arrows,arrows.meta,spy,matrix}
\usepgfplotslibrary{patchplots,groupplots,colorbrewer}
\usepackage{tikzscale}

\newif\ifexttikz
\exttikzfalse
\ifexttikz
	\usetikzlibrary{external}
\fi

\title{IEEE 802.11bf: Toward Ubiquitous Wi-Fi Sensing}

\author{\IEEEauthorblockN{Francesco Restuccia}\\
\IEEEauthorblockA{Department of Electrical and Computer Engineering, Northeastern University, US}
\vspace*{-1cm}
}

\usepackage{multirow}

\usepackage[font=footnotesize]{subcaption}
\usepackage[font=footnotesize]{caption}

\usepackage{mathtools}

\newacronym{6g}{6G}{sixth generation}
\newacronym{5g}{5G}{5th generation}

\newacronym{dcm}{DCM}{distributed cooperative \gls{mimo}}
\newacronym{comp}{CoMP}{Coordinated Multi-Point}

\newacronym{ssm}{SSM}{spectrum sensing module}
\newacronym{dsa}{DSA}{dynamic spectrum access}
\newacronym{bpd}{BPM}{baseband processing module}
\newacronym{drl}{DRL}{deep reinforcement learning}
\newacronym{5gb}{5GB}{5G and beyond}
\newacronym{ml}{ML}{machine learning}
\newacronym{cbrs}{CBRS}{Citizens Broadband Radio Service}
\newacronym{gaa}{GAA}{General Authorized Access}
\newacronym{pal}{PAL}{Priority Access Licensee}
\newacronym{fcc}{FCC}{Federal Communications Commission}
\newacronym{rfp}{RFP}{radio fingerprinting}
\newacronym{dsp}{DSP}{digital signal processing}
\newacronym{ssa}{SSA}{spectrum sensing and access}
\newacronym{fir}{FIR}{finite impulse response}
\newacronym{wsc}{WSC}{wireless signal classification}
\newacronym{ber}{BER}{bit error rate}
\newacronym{wwa}{WWI}{wideband waveform identification}
\newacronym{ppi}{PPI}{protocol and parameter inference}
\newacronym{ism}{ISM}{industrial, scientific and medical}
\newacronym{itu}{ITU}{international telecommunication union}
\newacronym{aml}{AML}{adversarial machine learning}
\newacronym{fml}{FML}{federated machine learning}
\newacronym{tml}{TML}{transfer machine learning}
\newacronym{api}{API}{Application Programming Interface}
\newacronym{mmw}{mmWave}{millimeter wave}

\newacronym{snr}{SNR}{Signal-to-Noise Ratio}
\newacronym{rssi}{RSSI}{Received Signal Strength Indicator}
\newacronym{cnn}{CNN}{convolutional neural network}
\newacronym{sp}{S\&P}{security and privacy}

\newacronym{wg}{WG}{Working Group}
\newacronym{tg}{TG}{Task Group}
\newacronym{par}{PAR}{Project Authorization Request}
\newacronym{frd}{FRD}{Functional Requirement Document}
\newacronym{bf}{TGbf}{IEEE 802.11bf}
\newacronym{ap}{AP}{Access Point}
\newacronym{sta}{STA}{Station}
\newacronym{sc}{SC}{Standards Committee}
\newacronym{ec}{EC}{Executive Committee}
\newacronym{sfd}{SFD}{Specification Framework Document}
\newacronym{csi}{CSI}{Channel State Information}
\newacronym{mac}{MAC}{Medium Access Control}
\newacronym{phy}{PHY}{Physical Layer}
\newacronym{ppdu}{PPDU}{\gls{phy} Protocol Data Unit}
\newacronym{mu}{MU}{Multiple User}
\newacronym{ndp}{NDP}{Null Data Packet}
\newacronym{mimo}{MIMO}{Multiple-Input Multiple-Output}
\newacronym{mumimo}{MU-MIMO}{\gls{mu} - \gls{mimo}}
\tikzstyle{startstop} = [rectangle, rounded corners, minimum width=2cm, minimum height=0.5cm,text centered, draw=black]
\tikzstyle{io} = [trapezium, trapezium left angle=70, trapezium right angle=110, minimum width=3cm, minimum height=1cm, text centered, draw=black]
\tikzstyle{process} = [rectangle, minimum width=2cm, minimum height=0.5cm, text centered, draw=black, alignb=center]
\tikzstyle{decision} = [ellipse, minimum width=2cm, minimum height=1cm, text centered, draw=black]
\tikzstyle{arrow} = [thick,<->,>=stealth]
\tikzstyle{line} = [thick,>=stealth]
\tikzstyle{darrow} = [thick,<->,>=stealth,dashed]
\tikzstyle{sarrow} = [thick,->,>=stealth]
\tikzstyle{larrow} = [line width=0.1mm,dashdotted,->,>=stealth]

\makeatletter
\def\grd@save@target#1{%
  \def\grd@target{#1}}
\def\grd@save@start#1{%
  \def\grd@start{#1}}
\tikzset{
  grid with coordinates/.style={
    to path={%
      \pgfextra{%
        \edef\grd@@target{(\tikztotarget)}%
        \tikz@scan@one@point\grd@save@target\grd@@target\relax
        \edef\grd@@start{(\tikztostart)}%
        \tikz@scan@one@point\grd@save@start\grd@@start\relax
        \draw[minor help lines] (\tikztostart) grid (\tikztotarget);
        \draw[major help lines] (\tikztostart) grid (\tikztotarget);
        \grd@start
        \pgfmathsetmacro{\grd@xa}{\the\pgf@x/1cm}
        \pgfmathsetmacro{\grd@ya}{\the\pgf@y/1cm}
        \grd@target
        \pgfmathsetmacro{\grd@xb}{\the\pgf@x/1cm}
        \pgfmathsetmacro{\grd@yb}{\the\pgf@y/1cm}
        \pgfmathsetmacro{\grd@xc}{\grd@xa + \pgfkeysvalueof{/tikz/grid with coordinates/major step x}}
        \pgfmathsetmacro{\grd@yc}{\grd@ya + \pgfkeysvalueof{/tikz/grid with coordinates/major step y}}
        \foreach \x in {\grd@xa,\grd@xc,...,\grd@xb}
        \node[anchor=north] at (\x,\grd@ya) {\pgfmathprintnumber{\x}};
        \foreach \y in {\grd@ya,\grd@yc,...,\grd@yb}
        \node[anchor=east] at (\grd@xa,\y) {\pgfmathprintnumber{\y}};
      }
    }
  },
  minor help lines/.style={
    help lines,
    gray,
    line cap =round,
    xstep=\pgfkeysvalueof{/tikz/grid with coordinates/minor step x},
    ystep=\pgfkeysvalueof{/tikz/grid with coordinates/minor step y}
  },
  major help lines/.style={
    help lines,
    line cap =round,
    line width=\pgfkeysvalueof{/tikz/grid with coordinates/major line width},
    xstep=\pgfkeysvalueof{/tikz/grid with coordinates/major step x},
    ystep=\pgfkeysvalueof{/tikz/grid with coordinates/major step y}
  },
  grid with coordinates/.cd,
  minor step x/.initial=.5,
  minor step y/.initial=.2,
  major step x/.initial=1,
  major step y/.initial=1,
  major line width/.initial=1pt,
}
\makeatother

\definecolor{desireRed}{RGB}{230,57,60}%
\definecolor{darkPurple}{RGB}{59,31,43}%
\definecolor{springGreen}{RGB}{37,223,145}%
\definecolor{queenBlue}{RGB}{69,123,157}%
\definecolor{spaceCadet}{RGB}{29,53,87}%

\makeglossaries

\begin{document}
\maketitle
% \thispagestyle{firstpage}
% \pagestyle{plain}

% Challenges: data fusion, intereference management/investigation

\begin{abstract}
Wi-Fi is among the most successful wireless technologies ever invented. As Wi-Fi becomes more and more present in public and private spaces, it becomes natural to leverage its ubiquitousness to implement groundbreaking wireless sensing applications such as human presence detection, activity recognition, and object tracking, just to name a few. This paper reports ongoing efforts by the IEEE 802.11bf Task Group (TGbf), which is defining the appropriate modifications to existing Wi-Fi standards to enhance sensing capabilities through 802.11-compliant waveforms. We summarize objectives and timeline of TGbf, and discuss some of the most interesting proposed technical features discussed so far. We also introduce a roadmap of  research challenges pertaining to Wi-Fi sensing and its integration with future Wi-Fi technologies and emerging spectrum bands, hoping to elicit further activities by both the research community and TGbf.
 \vspace{-0.5cm}
\end{abstract}

\glsresetall

\section{Introduction}

Since its inception in September 1990, Wi-Fi has transitioned from a low-rate replacement to Ethernet to one of the most successful wireless technologies ever invented. Today, Wi-Fi is ubiquitous and widely employed to provide plug-and-play Internet connectivity in almost any public and private space, including homes, offices, parks, airports, shopping malls, university campuses, and so on. Wi-Fi hotspots are so omnipresent that cellular operators are expected to offload 63\% of their global mobile data traffic to Wi-Fi by 2021. For this reason, it is not hard to believe that a whopping 541.6M public Wi-Fi hotspots will be deployed by 2021 -- a sixfold increase from 94M hotspots deployed in 2016, according to the Cisco Visual Networking Index (VNI) \cite{CiscoVNI}. As a consequence, the global Wi-Fi economy is forecast to increase from today's \$1.96T to \$3.46T by 2023, according to the study in \cite{WiFiAlliance}.

The staggering increase in Wi-Fi devices, coupled with the unprecedented throughput demands of next-generation multimedia applications, will inevitably create complex wireless networking challenges. However,  from a \textit{sensing} standpoint, the sudden explosion in Wi-Fi devices may also constitute a ``blessing in disguise.'' Indeed, Wi-Fi devices are not only becoming more bandwidth-hungry, but also more and more diverse, spanning from personal computers, smartphones, televisions, tablets, sensors and many others. This extremely dense and heterogeneous concentration of WiFi devices -- placed in almost every corner of our indoor environments -- will create the perfect opportunity to continuously ``map'' the surrounding environment using Wi-Fi signals as sounding waveforms.

\begin{figure}[!h]
    \centering
    \includegraphics[width=0.95\columnwidth]{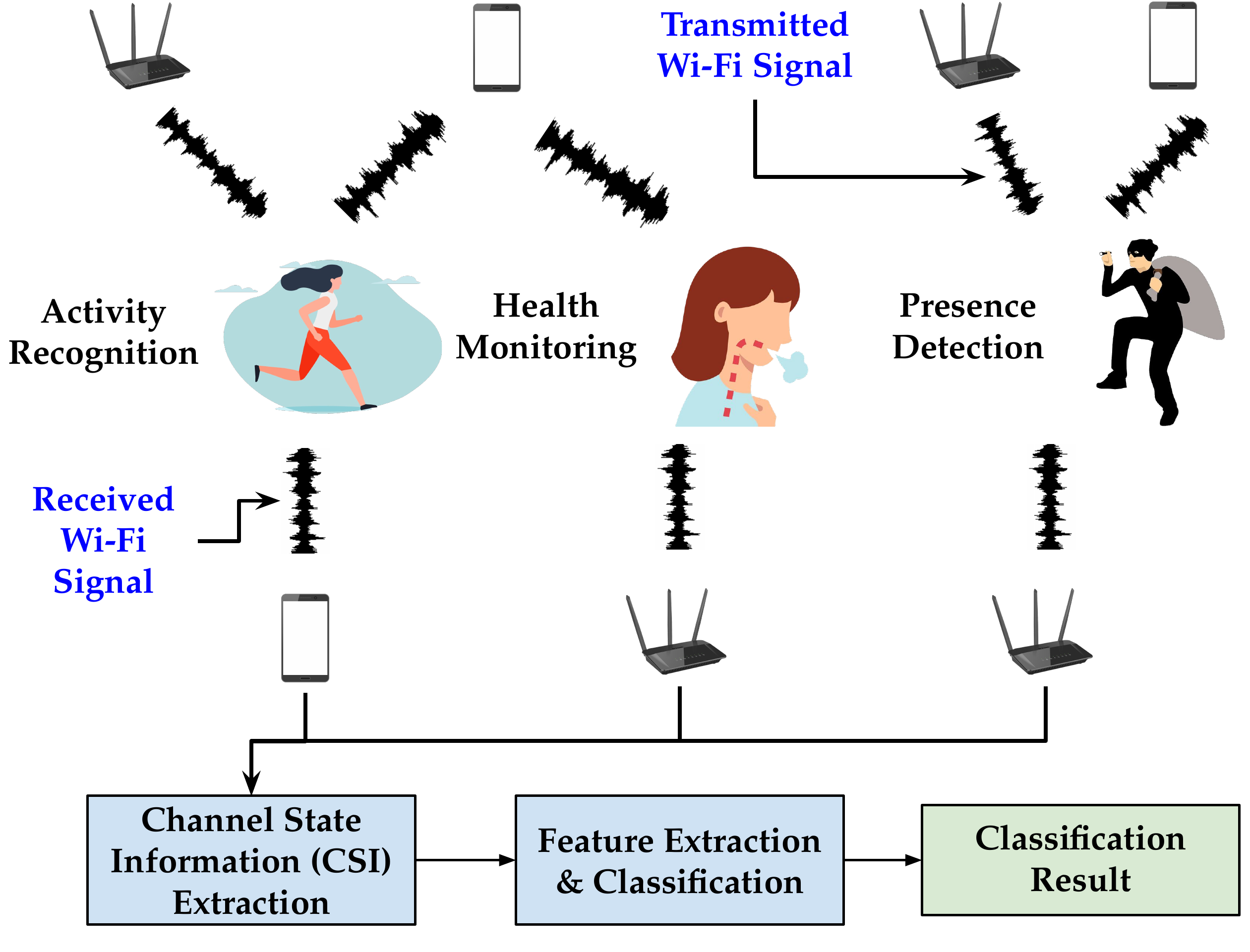}
    \caption{High-level Overview of Wi-Fi Sensing (SENS).}
    \label{fig:wifi_sensing}
\end{figure}

Figure \ref{fig:wifi_sensing} shows a very high-level overview of Wi-Fi sensing, also called SENS in the IEEE 802.11 community. The key idea behind SENS is to leverage \gls{csi} measurements to detect (and possibly, track) the presence of obstacles between a transmitter and a receiver. This way, complex classification tasks such as human activity recognition (HAR) \cite{ma2021location}, health monitoring \cite{wang2017tensorbeat}, and object detection \cite{zhu2017r} can be achieved, for example, through computation of phase differences \cite{wang2017phasebeat} or Doppler shifts \cite{qian2017widar}. Since surveying SENS procedures is outside the scope of this paper, we refer the reader to \cite{ma2019wifi} for an excellent summary.

The vast majority on the literature on \gls{csi}-based SENS has focused on finding new application scenarios, or improving the classification accuracy of existing SENS algorithms. This deluge of interest by the research community has definitely helped SENS establish itself as one of the top candidates to perform device-free sensing in the years to come. Somehow surprisingly, despite the success of SENS, current Wi-Fi standards do not provide the support for sensing activities. To truly transition SENS from a research-only process to a widespread, systematically used feature, extending existing Wi-Fi standards to support SENS become quintessential.  Recognizing the importance of the issue, the IEEE 802.11 \gls{wg} has approved in September 2020 a \gls{par} defining a new \gls{tg}, called \gls{bf} \cite{TGbfPAR}. According to the \gls{par}, \gls{bf} will tackle the development of an amendment that defines modifications to  state-of-the-art IEEE 802.11 standards at both the \gls{mac} and \gls{phy}. This will help enhance SENS operation in license-exempt frequency bands between 1 GHz and 7.125 GHz, as well as above 45 GHz, which will enable sensing at \gls{mmw} frequencies.

The impact that IEEE 802.11bf will have on our society at large cannot be overstated. When 802.11bf will be finalized and introduced as an IEEE standard in September 2024,  Wi-Fi will cease to be a communication-only standard and will legitimately become a \textit{full-fledged sensing paradigm}. This standardization effort, coupled with the numerous SENS systems currently being developed at the research stage, will create the ``perfect storm'' for the introduction into the market of groundbreaking applications that we cannot even imagine today. Through 802.11bf, Wi-Fi sensing will be merged into WiFi and become part of our daily lives, as every standard-compliant router will need to implement SENS capabilities. At the same time, IEEE 802.11bf is still in its infancy, and although the activities of \gls{bf} have only recently started and discussions are ongoing, some possible features that will become part of IEEE 802.11bf have already been discussed during \gls{bf}'s meetings. Moreover, there are still some critical issues that need to be addressed to really make SENS effective and efficient in a variety of circumstances. 

This paper makes the following contributions. First, we discuss in Section \ref{sec:need} a series of reasons as to why the standardization of SENS procedures through IEEE 802.11bf has become a compelling necessity for the future of Wi-Fi. We then introduce the reader to the use cases, objectives and timeline of \gls{bf} in Section \ref{sec:so_far}, and discuss the most relevant features that have been discussed during \gls{bf} meetings in Section \ref{sec:features}. We then introduce a series of research challenges that will need to be tackled to improve SENS and that could inform future \gls{bf} activities in Section \ref{sec:challenges}. The paper is concluded in Section \ref{sec:conclusions}.\vspace{-0.3cm}

\section{IEEE 802.11bf: The Need to Standardize Sensing Procedures}\label{sec:need}

Since its inception, SENS has enjoyed a sustained interest by the wireless research community, which has proposed and improved new techniques to leverage Wi-Fi signals to sense a diverse number of phenomena \cite{ma2019wifi}. After several years of continuous improvement, and realizing its  economic impact, the IEEE 802.11 community is now supporting the integration and facilitation of SENS procedures through the IEEE 802.11bf standard,  to increase compatibility, facilitate interoperability and support safety. This does not come as a surprise; to be successfully adopted at the scale of billions of devices, SENS techniques must necessarily be integrated into the bigger IEEE 802.11 family, which will impose manufacturers wanting to produce standard-compliant Wi-Fi chips to support SENS.

Beyond widespread adoption, we can provide at least two reasons that make 802.11bf compelling from a technical standpoint. First, the standardization process will impose strict quality control to the technical features that will be part of the standard. Promoting 802.11bf to a full standard will require several years and iterations, as explained in Section \ref{sec:timeline}, and will involve the participation of hundreds of experts in Wi-Fi technologies. Each proposed technical feature will go through many iterations before ending up into a standard draft. According to the \gls{bf} Selection Procedures document \cite{TGbfSP}, the proposed technical contributions will go through three voting procedures, where at least 75\% of the 802.11 voting members will have to approve the contributions. Finally, the draft document will need to satisfy 802.11bf technical requirements before making it to a \gls{wg} ballot.

The second reason is that through standardization into the 802.11 family, researchers and developers of sensing applications will have the access to well-defined SENS procedures, where data collection and analysis will be facilitated. Nowadays, researchers need to manually implement operations such as \gls{csi} collection, feature extraction, and classification. With established protocols defining the interaction with the \gls{phy} and \gls{mac} layers, innovation in the critical SENS field will be severely expedited.\vspace{-0.3cm}

\section{IEEE 802.11bf: \\Objectives and Timeline}\label{sec:so_far}

At the time of writing, the \gls{bf} community is regularly meeting 
to define the use cases and technical contribution that should be included in the final 802.11bf standard. However, a number of key objectives and use cases have already been defined, including the project timeline, which are the focus of this section.\vspace{-0.3cm}

% \textbf{Applications.}~The complete list -- still expanding -- of the IEEE 802.11bf use cases can be found at \cite{TGbfUC}. The most important are:

% \begin{itemize}
%     \item \textit{Room sensing}: human presence detection, people counting; 
%     \item \textit{Gesture recognition}: short range (finger movement), medium range (hand movement), large range (full body movement); 
%     \item \textit{Health care}: fall detection,  remote diagnostics, surveillance/monitoring of elder people and/or children;
%     \item \textit{Three-dimensional (3D) vision}: building a 3D picture of an environment;
%     \item \textit{In-car sensing}: detection of humans in car, driver sleepiness detection.
% \end{itemize}

\subsection{\gls{bf} Objectives}

The \gls{bf}'s final target is to define modifications to the 802.11ad, 802.11ay, 802.11n, 802.11ac, 802.11ax and 802.11be \gls{phy}, as well as the IEEE 802.11 \gls{mac}, to enhance sensing operations through Wi-Fi signals. Finally, 802.11bf will provide backward compatibility and coexistence with existing IEEE 802.11 standards operating in the same bands. It will enable sensing measurements to be obtained using transmissions that are requested, unsolicited, or both, and define a \gls{mac} service interface for layers above the \gls{mac} to request and retrieve SENS measurements.

More formally, the \gls{bf} defines Wi-Fi Sensing (SENS) as the usage of received Wi-Fi signals from a Wireless \gls{sta} to detect features (i.e., range, velocity, angular, motion, presence or proximity, gesture, etc) of intended targets (i.e., object, human, animal, etc) in a given environment (i.e., house, office, room, vehicle, enterprise, etc). As specified in the \gls{frd} \cite{TGbfFRD} redacted by the \gls{bf}, SENS operations will occur in license-exempt frequency bands between 1 GHz and 7.125 GHz and above 45 GHz. The IEEE 802.11bf amendment will enable \glspl{sta} to perform the following: (i) exchange information regarding their SENS capabilities to other \glspl{sta}; (ii) request and setup transmissions that enable SENS measurements to be performed; (iii) indicate a transmission can be used for SENS; (iv) exchange of SENS feedback and information.

\begin{figure}[!h]
    \centering
    \includegraphics[width=0.95\columnwidth]{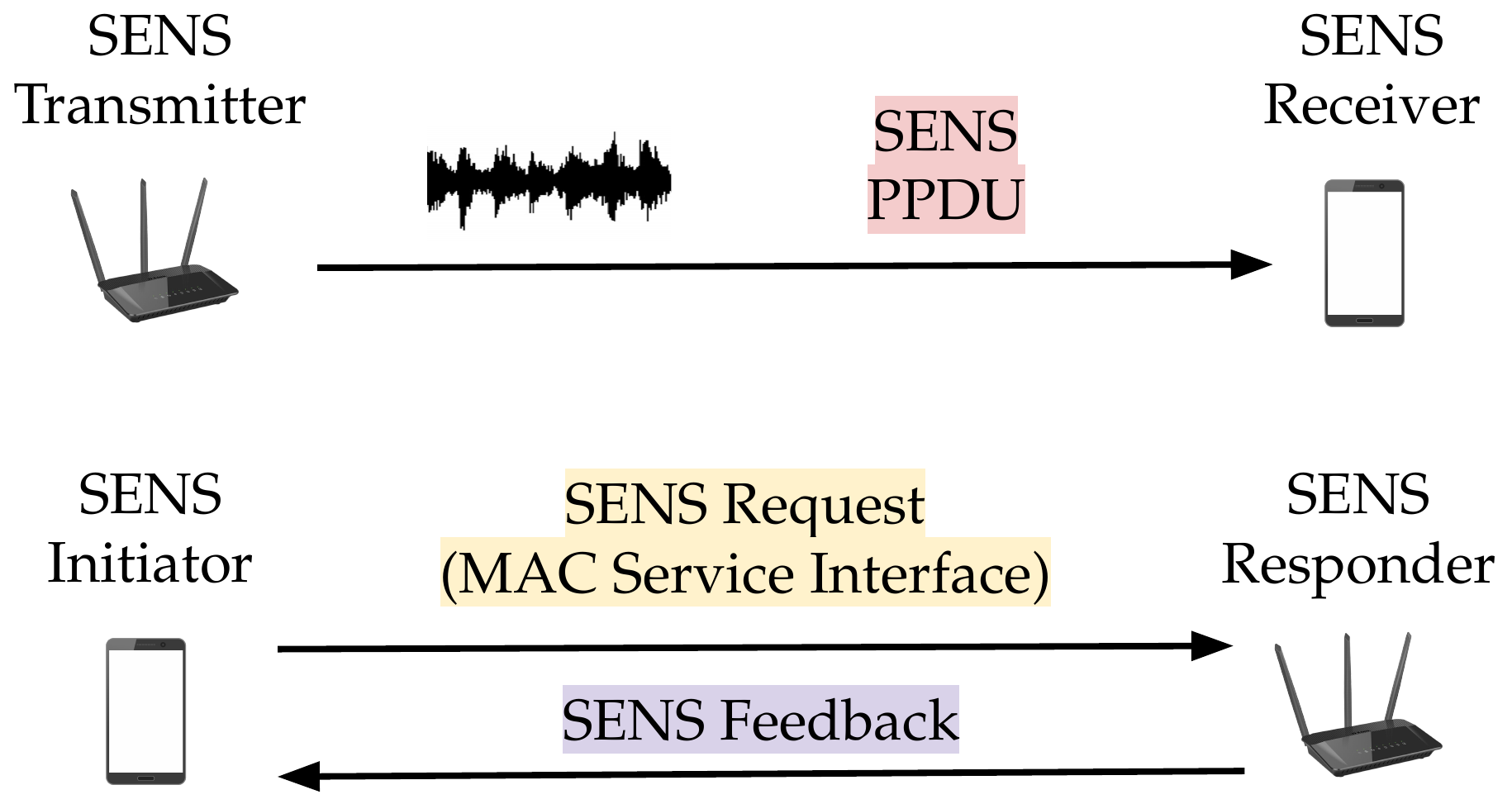}
    \caption{Support for SENS Operations in 802.11bf.\vspace{-0.4cm}}
    \label{fig:sens_ops}
\end{figure}

Figure \ref{fig:sens_ops} summarizes how 802.11bf is envisioned to  support sensing operations. SENS operations will be performed during a SENS \emph{session}. A SENS initiator is a \gls{sta} that initiates a session, while a SENS responder is a \gls{sta} that participates to a session started by an initiator. A SENS transmitter/receiver will transmit/receiver \glspl{ppdu} used for SENS measurements in a session. Notice that in a session, an initiator can be a sensing transmitter, a receiver, both or neither.  

\begin{figure*}
    \centering
    \includegraphics[width=\textwidth]{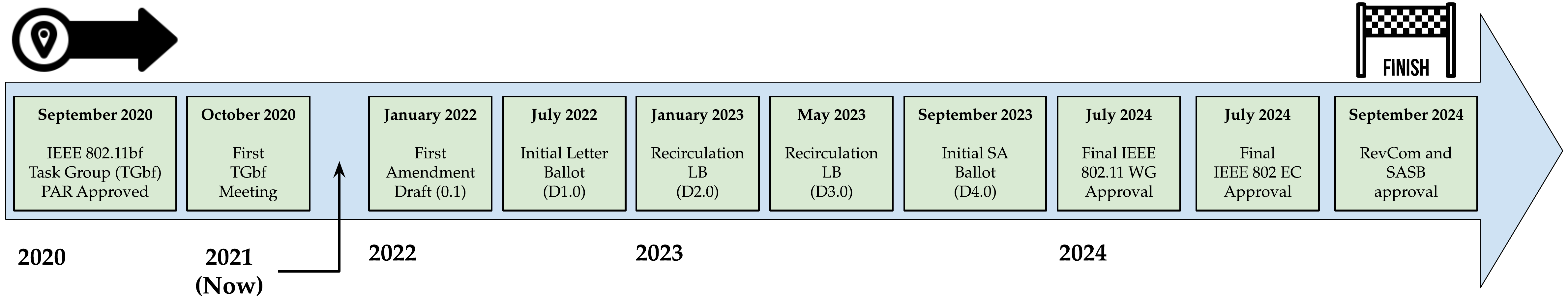}
    \caption{Development Timeline of the IEEE 802.11bf Standard.\vspace{-0.3cm}}
    \label{fig:bf_timeline}
\end{figure*}

\subsection{\gls{bf} Timeline}\label{sec:timeline}

Figure \ref{fig:bf_timeline} summarizes the envisioned 4-year development timeline for 802.11bf. The first crucial step, which is the approval of the Project Authorization Request (PAR) for \gls{bf}, has been completed in September 2020. In IEEE standards jargon, a \gls{par} is a legal document that states the reason and objectives for the project. Moreover, it allows a \gls{wg} to assign copyright and indemnification from IEEE. At the moment, \gls{bf} is working on drafting a \gls{sfd} to provide a top-level technical description of the functionality that the IEEE 802.11bf standard will implement. 

An initial draft of the standard (D0.1) is expected to be available in January 2022, with 3 subsequent drafts, D1.0, D2.0 and D3.0) available in July 2022, January 2023 and May 2023, respectively. The IEEE Standard Association (SA) balloting process for IEEE 802.11bf is set to begin in September 2023. The balloting usually begins when the \gls{sc} assigned to \gls{bf} convenes that a draft (D4.0) has reached enough maturity. A balloting group will be formed by the \gls{sc}, containing individuals or entities interested in the standard. While anyone can contribute comments, eligible members of the balloting group (i.e., IEEE SA members or buyers of a per-ballot fee) are the only votes that count toward approval. Balloters are usually stakeholders of the standard, such as chip manufacturers and final users. While no interest category can comprise over 1/3 of the balloting group, the target is to gain the greatest consensus among balloters. At least, a 75\% consensus with at least 75\% of the ballots returned has to be reached to conclude the balloting process. 

The IEEE 802.11 \gls{wg} and 802.11 \gls{ec} is set to approve the 802.11bf standard in July 2024, 10 months after the SA ballot. Before the standard can be published, the IEEE SA Standards Board has to approve it with the recommendation of the Standards Review Committee (RevCom). While the RevCom does not examine the technical nature of the standard, it ensures procedural standards were followed during the drafting and balloting process, such as consensus, due process, openness, and balance. After approval, the standard is edited by an IEEE SA editor, given a final review by the members of the \gls{wg}, and published. RevCom will check over all the documentation and make sure that the IEEE SA procedures were followed. \vspace{-0.3cm}

\section{IEEE 802.11bf: Proposed Features}\label{sec:features}

Many technical contributions are currently being discussed during \gls{bf} meetings to take SENS to the next level in future Wi-Fi networks. In this section, we provide a summary of the most promising approaches discussed so far.\vspace{-0.4cm}

\subsection{Cooperative SENS}\label{sec:csens}

\begin{figure}[!h]
    \centering
    \includegraphics[width=0.95\columnwidth]{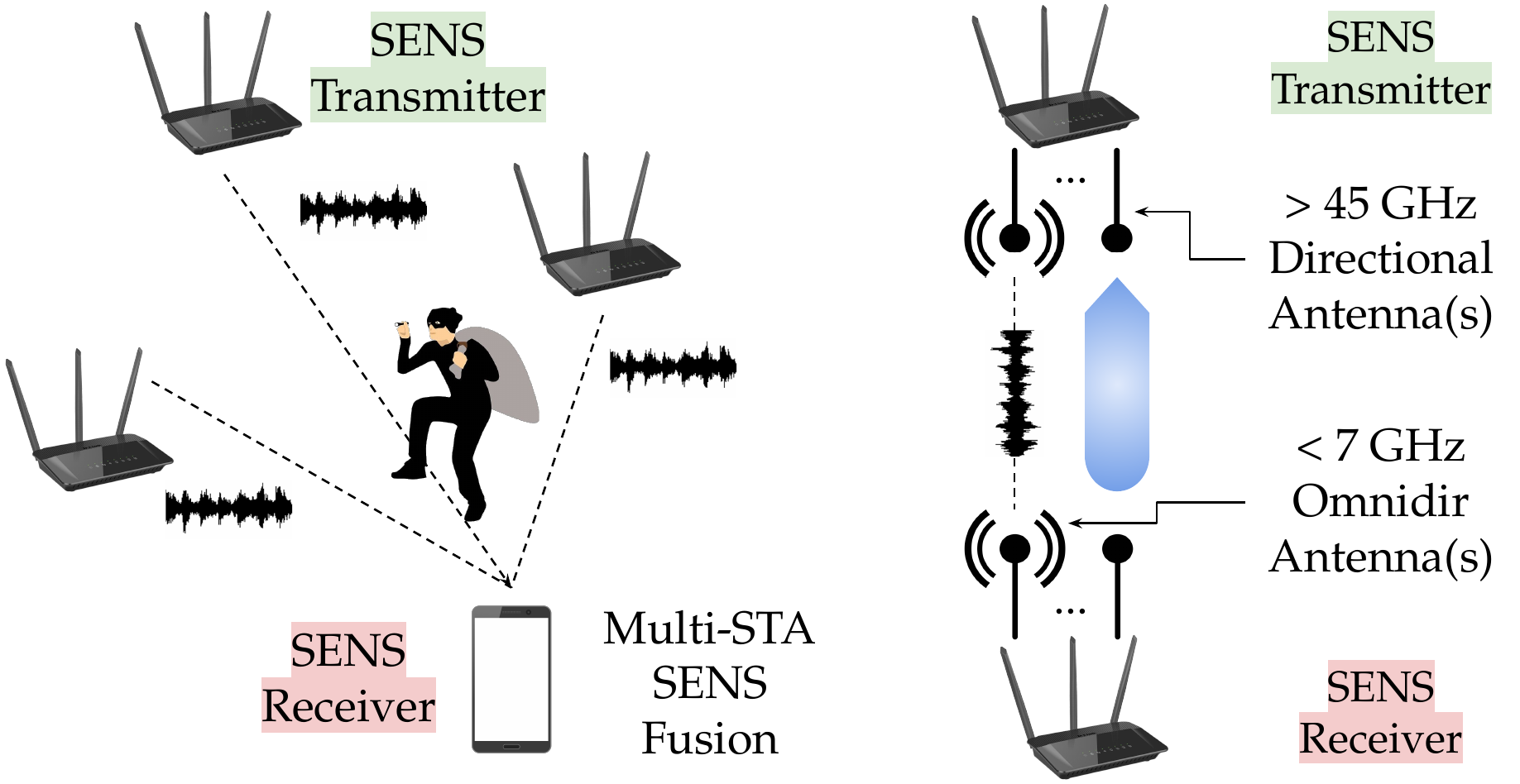}
    \caption{Cooperative and Multi-band SENS Approaches in 802.11bf.\vspace{-0.4cm}}
    \label{fig:sens_approaches}
\end{figure}

Traditionally, SENS has performed with one \gls{sta} transmiting the sensing waveform and one (or more) \glspl{sta} receiving and processing it. However, a more effective \emph{collaborative} SENS (in short, CSENS) approach has been recently proposed  \cite{CollabSensing}, shown in the left portion of Figure \ref{fig:sens_approaches}, where multiple SENS-enabled devices can collaborate as a group in an orderly fashion to capture additional information about the surrounding environment. 

\begin{figure}[!h]
    \centering
    \includegraphics[width=1\columnwidth]{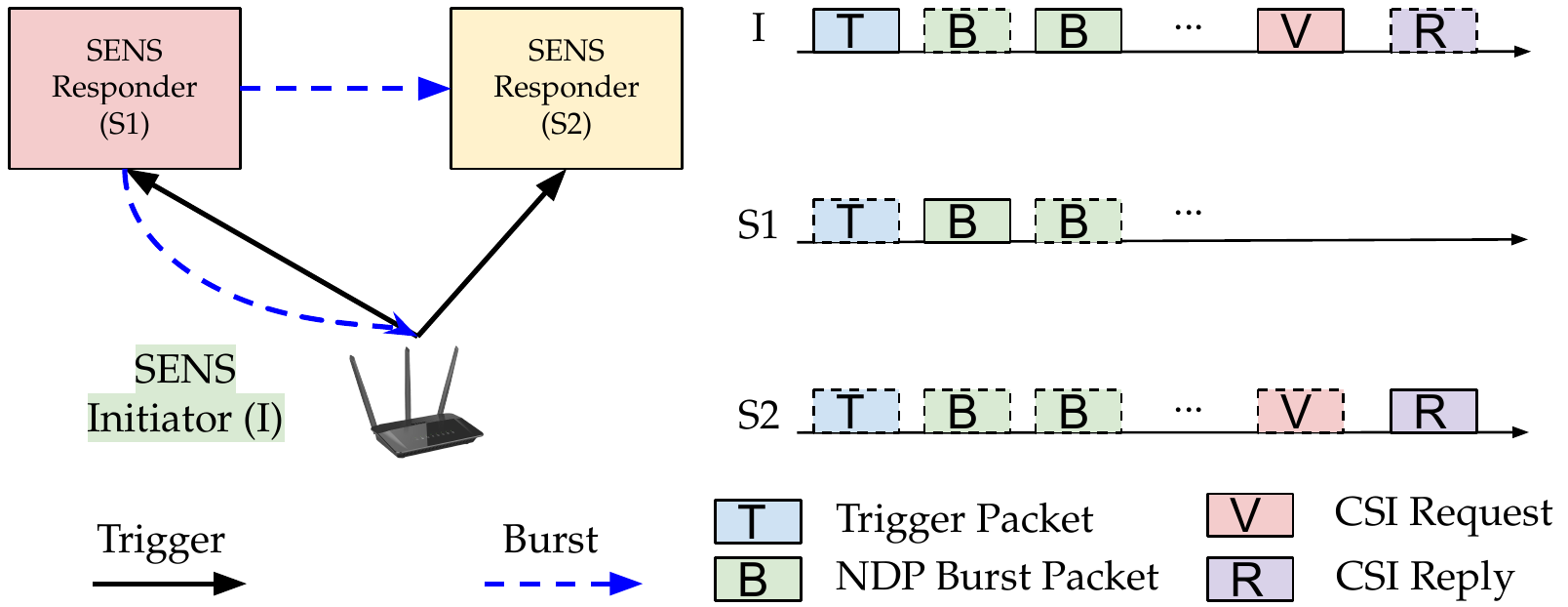}
    \caption{Example of a CSENS session.\vspace{-0.2cm}}
    \label{fig:sens_example}
\end{figure}

This is especially true in the case of \gls{mimo} transmissions, which are the norm in modern Wi-Fi standards. Spatial diversity -- achieved with multiple antennas and multiple transmitters -- can indeed help improve the quality of SENS operations, as more channels between the SENS transmitters and the SENS receiver(s) can be obtained and processed, thus leading to accurate sensing. Figure \ref{fig:sens_example} illustrates  the envisioned way to establish a CSENS session, which is composed of three phases: (i) a trigger phase, (ii) a burst phase, and (iii) a feedback exchange phase. A SENS initiator (I) initiates the CSENS by sending a trigger packet, which signals the SENS responders (S1 and S2) that a CSENSE operation is taking place. The responders will take turn in transmitting a burst of \glspl{ndp}, which are used to sound the channel between the responders and the receiving \glspl{sta}. 

In Figure \ref{fig:sens_example}, S1 sends a burst of two NDPs, one received by I and one by S2. Since an NDP is received by I, there is no need to request an explicit feedback since I can directly compute the \gls{csi}. To receive the \gls{csi} from the missing responders, I sends a feedback request -- in the example, S2 -- each of them replying with a  \gls{csi} reply.\vspace{-0.3cm}

\subsection{Multi-band SENS}\label{sec:multi_band}

The vast majority of Wi-Fi devices will soon be equipped with antennas operating at both sub-7-GHz frequencies and 60 GHz, to be compatible with legacy standards -- IEEE 802.11ax and earlier -- and the new standards at  \gls{mmw} frequencies -- IEEE 802.11ad/ay and subsequent. The presence of such diverse antennas into a single device will create unprecedented sensing opportunity for Wi-Fi. On one hand, sub-7 \gls{csi} measurements will provide indication of relatively large motions, can propagate through obstacles (e.g., walls), and contain richer multipath information. On the other hand, \gls{csi} measurements are very sensitive to fading and noise, which may lead to inconsistencies in sensing measurements. \gls{csi} is also very high-dimensional, as it grows quadratically with the number of  subcarriers and antennas. Conversely, \gls{rssi} measurements at \gls{mmw} will provide highly-directional information through the usage of beamforming toward a given receiver, but have small range due to the presence of blockers (e.g., walls). Moreover, measurement are more coarse-grained, since the complexity grows with the number of beams. 

Given the substantial difference between the two measurements, it becomes consequential to ``merge'' together sensing inputs from sub-7 and \gls{mmw}. This is particularly useful especially for data-driven algorithms, which can work with very heterogeneous data \cite{restuccia2020deep}. Recent discussions in \gls{bf} meetings \cite{MultiBand} have unveiled the possibility of concurrently using \gls{rssi} and \gls{csi} measurements as input to a \gls{cnn} or \gls{dsp} block to classify complex sensing phenomena. The approaches discussed were input concatenation, feature fusion, i.e., concatenate the output of the convolutional layers, and feature permutation, to make the learning process more robust and generalizable.\vspace{-0.3cm}

\section{IEEE 802.11bf: \\Challenges and Opportunities}\label{sec:challenges}

The IEEE 802.11bf standardization process is still in its infancy. Although significant research efforts have been spent in the field of Wi-Fi sensing, there are still a plethora of opportunities to further enrich the field before the standardization process is completed in September 2024. Figure \ref{fig:sens_challenges} summarizes the challenges discussed in this section.\vspace{-0.4cm}

\subsection{SENS Security and Privacy}

While creating a plethora of life-improving benefits for ordinary citizens, the IEEE 802.11bf standard will enable Wi-Fi devices to regularly perform SENS operations in highly-populated indoor environments. As a consequence, the pervasiveness of SENS into our everyday lives will necessarily elicit \gls{sp} concerns by the end users. Indeed, it has been shown that SENS-based classifiers can infer privacy-critical information such as keyboard typing, gesture recognition and activity tracking. Given the broadcast nature of the wireless channel, a malicious eavesdropper could easily ``listen'' to \gls{csi} reports and track the user's activity without authorization. Worse yet, since Wi-Fi signals can penetrate hard objects and can be used without the presence of light, end-users may not even realize they are being tracked. 

As yet, research and development efforts have been focused on improving the classification accuracy of the phenomena being monitored, with little regard to \gls{sp} issues. While this could be acceptable from a research perspective, we point out that to allow widespread adoption of 802.11bf, ordinary people need to trust its underlying technologies. Therefore, \gls{sp} guarantees must be provided to the end users.

\begin{figure*}
    \centering
    \includegraphics[width=\textwidth]{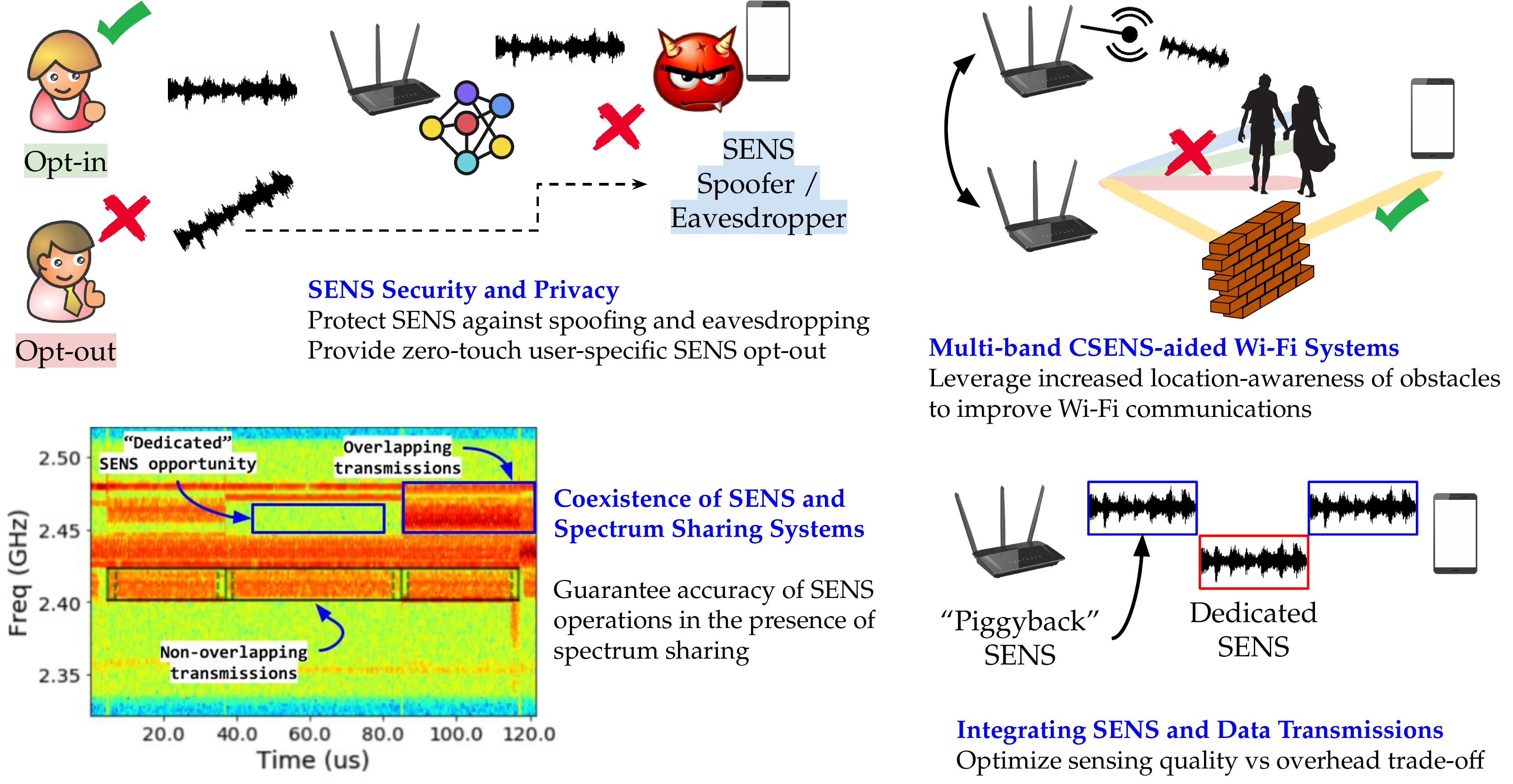}
    \caption{Challenges and Opportunities in Wi-Fi Sensing.\vspace{-0.4cm}}
    \label{fig:sens_challenges}
\end{figure*}

We identify a number of critical issues that need to be addressed in this space, also illustrated in the top-left portion of Figure \ref{fig:sens_challenges}. First, individuals should be provided the opportunity to \textit{opt out} of SENS services -- in other words, to avoid being monitored and tracked by the Wi-Fi devices around them. This would require the widespread introduction of reliable SENS algorithm for human or animal identification. Although some techniques have been proposed in literature \cite{ma2019wifi}, it is unclear whether they are resilient to spoofing, i.e., malicious users actively trying to impersonate other users, or adverse channel conditions, i.e., presence of noise and interference from other technologies. On the other hand, identification techniques should also be tested against adversaries trying to avoid being detected, either through active techniques, i.e., a Wi-Fi device carefully jamming the SENS activity,  or passive techniques, i.e., materials shielding and/or deflecting the Wi-Fi radiation. Since many recent SENS-based classification systems are based on \glspl{cnn}, an interesting investigation could be the evaluation of the extent to which adversarial machine learning (AML) techniques can compromise the accuracy of existing \gls{cnn}-based classifiers operating on \gls{csi} inputs. \vspace{-0.3cm}

\subsection{Multi-band CSENS-aided Wi-Fi Systems}

Cooperative SENS (CSENS), as mentioned in Section \ref{sec:csens}, has been discussed as a viable option to increase the reliability of SENS operations. This feature,  combined with the possibility of multi-band SENS (Section \ref{sec:multi_band}), will provide a unique opportunity to not only increase the classification accuracy of sensed phenomena, but also to leverage the increased location-awareness of blockages -- due to humans, animals and objects -- to design intelligent multi-band CSENS-aided Wi-Fi communications that will increase the performance of \gls{mmw} Wi-Fi links. For example, understanding the size and movement of blocking entities through sub-7 \gls{csi} reports could eventually guide beam selection in the \gls{mmw} link, as shown in Figure \ref{fig:sens_challenges}. By the same token, understanding the location of a \gls{sta} by using sub-7 SENS can help reduce the overhead associated with beam scanning and alignment. 

Moving forward, a key challenge will be to coordinate time-sensitive CSENS operations among multiple Wi-Fi devices in different spectrum bands. Indeed, conversely from the vast majority of SENS classification tasks, communication-related SENS will be extremely time-sensitive, with maximum tolerable deadlines in the order of milliseconds. To this end, a possible strategy could be to introduce control channels in the sub-7 band exclusively dedicated to the coordination of low-latency CSENS operations. This option is particularly enticing, also thanks to the increased number -- up to 16 -- of \gls{mimo} spatial streams supported by future Wi-Fi standards such as IEEE 802.11be \cite{lopez2019ieee}. \vspace{-0.3cm}

\subsection{SENS in Spectrum Sharing Environments}

From IEEE 802.11ax onward, Wi-Fi is poised to utilize the additional 1.2 GHz of spectrum between 5.925 and 7.125 GHz. Of particular relevance to IEEE 802.11bf, this pristine spectrum band will allow the usage of very large bandwidths -- up to 320 MHz \cite{lopez2019ieee} -- and as a consequence, the usage of an increased number of subcarriers -- 996x2, if the same subcarrier allocation of IEEE 802.11ax will persist. This will imply that finer-grained \gls{csi} reports could be utilized for SENS, with the possibility of severely improving  performance.

Two major showstoppers to this much-need enhancement, however, are (i) the increased path loss at 6 GHz frequencies; and (ii) the need to share the spectrum with other wireless technologies. Currently, the 6 GHz band is reserved to licensed users (also called incumbents). These include cellular carriers and mobile virtual network operators (MVNOs) who have already deployed thousands of backhaul point-to-point links. Incumbents such as satellite, public safety and ultra-wide bandwidth systems will also be in the 6 GHz band, as well as upcoming \gls{5g} technologies such as NR-Unlicensed. To protect incumbent services, restrictions will likely be placed in terms of maximum emitted power and communication time by other technologies. Thus, (i) Wi-Fi transmissions may utilize significantly lower power in some cases, and (ii) severe interference from incumbent and upcoming SENS has to be expected, especially when listen-before-talk is not used to handle coexistence. Therefore, further investigations should  address how to make SENS robust to  interference.%\vspace{-0.4cm}

\subsection{Integrating SENS and Data Transmissions}

SENS operations are set to coexist with data-only transmissions in future 802.11 standards. On one hand, since both data packets (DPs) and \glspl{ndp} contain \gls{csi}, SENS transmissions could be ``piggybacked'' into DPs to avoid decreasing throughput to a significant extent. On the other hand, as explained earlier, DPs may be subject to significant interference in the 6 GHz band, which may be tolerable from a data recovery perspective but intolerable from a SENS perspective. Therefore, a core issue is to determine the optimal trade-off between making reserved use of the spectrum for SENS operations and piggybacking SENS into DP. Similar to multi-band SENS, dedicated channels -- either through \gls{mimo} spatial multiplexing or through periodic channel reservation -- could be used to improve SENS performance without a significant decrease in system throughput.

\section{Conclusions} \label{sec:conclusions}

With the introduction of IEEE 802.11bf, we are starting to see almost 10 years of research efforts come to fruition. However, there are still many challenges and opportunities for the research community to contribute to a successful standard. In this paper, we have summarized current IEEE 802.11bf standardization efforts, as well as some of the proposed technical features. We have also discussed a set of research challenges that if addressed, could improve the finalized IEEE 802.11bf standard. We hope our paper will elicit further research activities by the wireless research community on this timely topic.

% \hspace{-0.2cm}
\footnotesize
\bibliographystyle{IEEEtran}
\bibliography{bibl}

% Generated by IEEEtran.bst, version: 1.14 (2015/08/26)
\begin{thebibliography}{10}
\providecommand{\url}[1]{#1}
\csname url@samestyle\endcsname
\providecommand{\newblock}{\relax}
\providecommand{\bibinfo}[2]{#2}
\providecommand{\BIBentrySTDinterwordspacing}{\spaceskip=0pt\relax}
\providecommand{\BIBentryALTinterwordstretchfactor}{4}
\providecommand{\BIBentryALTinterwordspacing}{\spaceskip=\fontdimen2\font plus
\BIBentryALTinterwordstretchfactor\fontdimen3\font minus
  \fontdimen4\font\relax}
\providecommand{\BIBforeignlanguage}[2]{{%
\expandafter\ifx\csname l@#1\endcsname\relax
\typeout{** WARNING: IEEEtran.bst: No hyphenation pattern has been}%
\typeout{** loaded for the language `#1'. Using the pattern for}%
\typeout{** the default language instead.}%
\else
\language=\csname l@#1\endcsname
\fi
#2}}
\providecommand{\BIBdecl}{\relax}
\BIBdecl

\bibitem{CiscoVNI}
{Cisco Inc.}, ``{Cisco Visual Networking Index (VNI) and VNI Service Adoption
  Global Forecast Update, 2016–2021},''
  \url{https://tinyurl.com/CiscoVNI2021}, 2020.

\bibitem{WiFiAlliance}
{Wi-Fi Alliance}, ``{The Economic Value of Wi-Fi: A Global View (2018 and
  2023)},'' \url{https://tinyurl.com/EconWiFi}, 2021.

\bibitem{ma2021location}
Y.~Ma, S.~Arshad, S.~Muniraju, E.~Torkildson, E.~Rantala, K.~Doppler, and
  G.~Zhou, ``{Location-and Person-Independent Activity Recognition with WiFi,
  Deep Neural Networks, and Reinforcement Learning},'' \emph{ACM Transactions
  on Internet of Things}, vol.~2, no.~1, pp. 1--25, 2021.

\bibitem{wang2017tensorbeat}
X.~Wang, C.~Yang, and S.~Mao, ``{TensorBeat: Tensor Decomposition for
  Monitoring Multiperson Breathing Beats with Commodity WiFi},'' \emph{ACM
  Transactions on Intelligent Systems and Technology (TIST)}, vol.~9, no.~1,
  pp. 1--27, 2017.

\bibitem{zhu2017r}
H.~Zhu, F.~Xiao, L.~Sun, R.~Wang, and P.~Yang, ``{R-TTWD: Robust device-free
  through-the-wall detection of moving human with WiFi},'' \emph{IEEE Journal
  on Selected Areas in Communications}, vol.~35, no.~5, pp. 1090--1103, 2017.

\bibitem{wang2017phasebeat}
X.~Wang, C.~Yang, and S.~Mao, ``{PhaseBeat: Exploiting CSI Phase Data for Vital
  Sign Monitoring with Commodity WiFi Devices},'' in \emph{Proc. of IEEE
  ICDCS}, 2017.

\bibitem{qian2017widar}
K.~Qian, C.~Wu, Z.~Yang, Y.~Liu, and K.~Jamieson, ``{Widar: Decimeter-level
  Passive Tracking via Velocity Monitoring with Commodity Wi-Fi},'' in
  \emph{Proc. of ACM MobiCom}, 2017.

\bibitem{ma2019wifi}
Y.~Ma, G.~Zhou, and S.~Wang, ``{WiFi Sensing with Channel State Information: A
  Survey},'' \emph{ACM Computing Surveys (CSUR)}, vol.~52, no.~3, pp. 1--36,
  2019.

\bibitem{TGbfPAR}
{IEEE 802.11bf Task Group (TG)}, ``{IEEE 802.11bf (TGbf) Project Authorization
  Request (PAR)},'' \url{https://tinyurl.com/TGbfPAR}, 2021.

\bibitem{TGbfSP}
------, ``{IEEE 802.11bf (TGbf) Selection Procedure Document},''
  \url{https://tinyurl.com/TGbfSP}, 2021.

\bibitem{TGbfFRD}
------, ``{IEEE 802.11bf (TGbf) Functional Requirements Document (FRD)},''
  \url{https://tinyurl.com/TGbfFRD}, 2021.

\bibitem{CollabSensing}
S.~Kim, D.~Lim, I.~Jang, J.~Kim, and J.~Choi, ``{Collaborative WLAN Sensing --
  Follow Up},'' \url{https://tinyurl.com/CollabSensingTGbf}, 2021.

\bibitem{restuccia2020deep}
F.~Restuccia and T.~Melodia, ``{Deep Learning at the Physical Layer: System
  Challenges and Applications to 5G and Beyond},'' \emph{IEEE Communications
  Magazine}, vol.~58, no.~10, pp. 58--64, 2020.

\bibitem{MultiBand}
P.~Wang, J.~Yu, T.~Koike-Akino, Y.~Wang, and P.~V. Orlik, ``{Multi-Band WiFi
  Fusion for WLAN Sensing},'' \url{https://tinyurl.com/TGbfMultiBand}, 2021.

\bibitem{lopez2019ieee}
D.~L{\'o}pez-P{\'e}rez, A.~Garcia-Rodriguez, L.~Galati-Giordano, M.~Kasslin,
  and K.~Doppler, ``{IEEE 802.11be Extremely High Throughput: The Next
  Generation of Wi-Fi Technology Beyond 802.11ax},'' \emph{IEEE Communications
  Magazine}, vol.~57, no.~9, pp. 113--119, 2019.

\end{thebibliography}

\begin{IEEEbiographynophoto}
{Francesco Restuccia} (M'16) is an Assistant Professor of Electrical and Computer Engineering at Northeastern University, United States. His research interests lie in embedded systems, wireless networks and artificial intelligence. 
\end{IEEEbiographynophoto}

\end{document}